# Two Paths to Learning Physics: How Games and Simulations Shape Physics Learning Among Physics and Engineering Students

Koushiki Pohit

Department of Technology – Purdue Polytechnic, Purdue University, West Lafayette, IN, 47907, U.S.A.

Razan Hamed, N. Sanjay Rebello

Department of Physics and Astronomy, Purdue University, West Lafayette, IN, 47907, U.S.A.

The use of digital tools like educational video games and interactive simulations is of great importance to physics education. This study investigates the sequencing effects of an educational game, *'Photon Jump'* and *PhET Photoelectric Effect* simulation on students' perception of such tools for learning about the photoelectric effect. Using a counterbalanced quasi-experimental crossover design, a total of 55 physics and engineering students from a calculus-based physics course were divided into two groups of comparable sizes and administered the game and the simulation in reverse order followed by 9 open ended reflection questions. The results show a clear preference for the game-based activity over the simulation. Additionally, gaming frequency showed no correlation with willingness to use similar tools, suggesting broad accessibility. Thematic analysis revealed that intuitive and explorative learning through the game was reinforced by the analytical aspect of the simulation.

## 1. INTRODUCTION & BACKGROUND

Educational video games and interactive simulations have gained increasing traction as digital tools to enhance conceptual learning in science education. Physics presents unique opportunities for innovative digital instruction due to its abstract and often counterintuitive concepts. The photoelectric effect, a phenomenon foundational to quantum physics, is one such concept that students frequently find challenging. Traditional instruction often relies on a single representational format, yet a large body of learning science shows that coordinating multiple, well-designed representations improves comprehension and transfer of learning especially for complex science topics (Ainsworth, 2006). Simulations provide interactive visualizations to replicate physical systems with a high degree of fidelity, helping students explore abstract scenarios (Dejong, 1998; Rutten, 2012). Well-known platforms like PhET Simulations have demonstrated consistent success in supporting inquiry-based learning, visualizing unseen forces, and facilitating exploration of complex relationships such as velocity, acceleration, and energy transfer (Perkins, 2006). Research shows that simulations can be particularly effective in correcting misconceptions when coupled with guided inquiry (Zacharia, 2011; Adams, 2008).

In contrast, educational games embed learning within narrative, challenge-based contexts. Such games represent physics or other types of phenomena analogically through metaphors, allegories, or playful constraints that do not mirror exact physical systems but invite conceptual transfer instead (Gee, 2005; Clark, 2016). Games are usually praised for their motivational appeal, capacity for sustained engagement, and alignment with constructivist learning theories (Squire, 2013; Papastergiou, 2009). They promote situated learning in which learners acquire knowledge through contextualized activity and feedback loops embedded within a game system (Barab, 2009). According to Gentner's Structure-Mapping Theory, analogical reasoning enables learners to map relationships from a familiar or fictional domain onto a target concept, fostering deeper understanding through metaphorical connections (Gentner, 1983).

Several studies (Mayer, 1992; Rieber, 1996) highlight the complementary nature of these approaches. Metaphor-based representations in games can enhance initial understanding, while simulations refine and stabilize that understanding through detailed feedback. Nevertheless, no known studies have explored the effect of sequencing of games and simulations on students' perception and science learning. Therefore, this study investigates: **RQ1) How do students perceive the use of a video game as a learning tool? RQ2) How does the sequence of instructional interventions (simulation followed by game vs. game followed by simulation) affect students' preferences and experience with each tool?**

## 2. METHODS

The participants were 55 students taking a calculus-based introductory physics course at a large public Midwestern university. Students were randomly divided into two groups. Each group completed the same simulation- and game-based activities in reverse order to examine potential order effects (Fig. 1). All students worked individually and completed a post-survey.

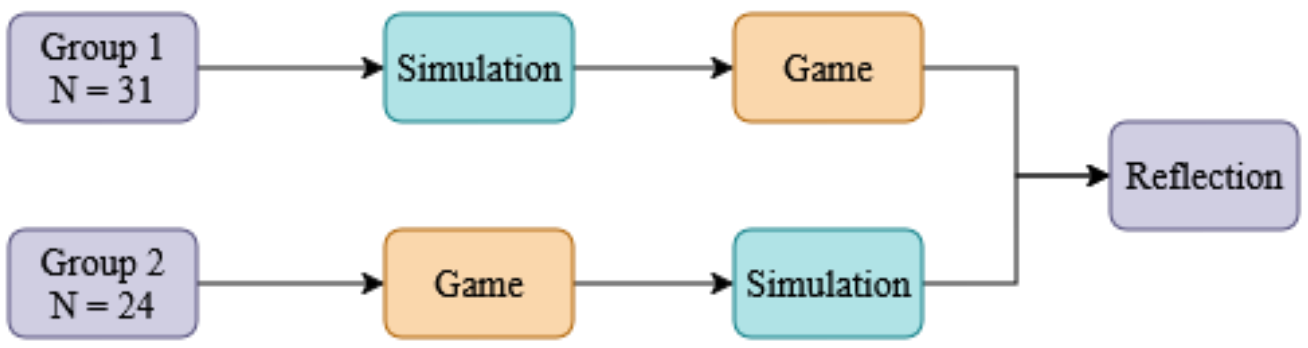

Figure 1. Experiment Design

**Photon Jump Game:** Students played Photon Jump (Fig. 2), 3D Unity-based game designed to model the photoelectric effect. Players take the role of an electron confined in a potential well. The aim is to gain energy from incoming photons and escape to light a bulb. A guided worksheet accompanied the game, prompting students to connect in-game elements with underlying physical concepts.

**PhET Photoelectric Effect Simulation:** Developed by PhET (2006) (Fig. 3), is a scientific tool which allows manipulation of light wavelength, intensity, and metal work function. A structured worksheet supported systematic exploration, data recording, and interpretation of observable photoelectron emission and energy graphs.

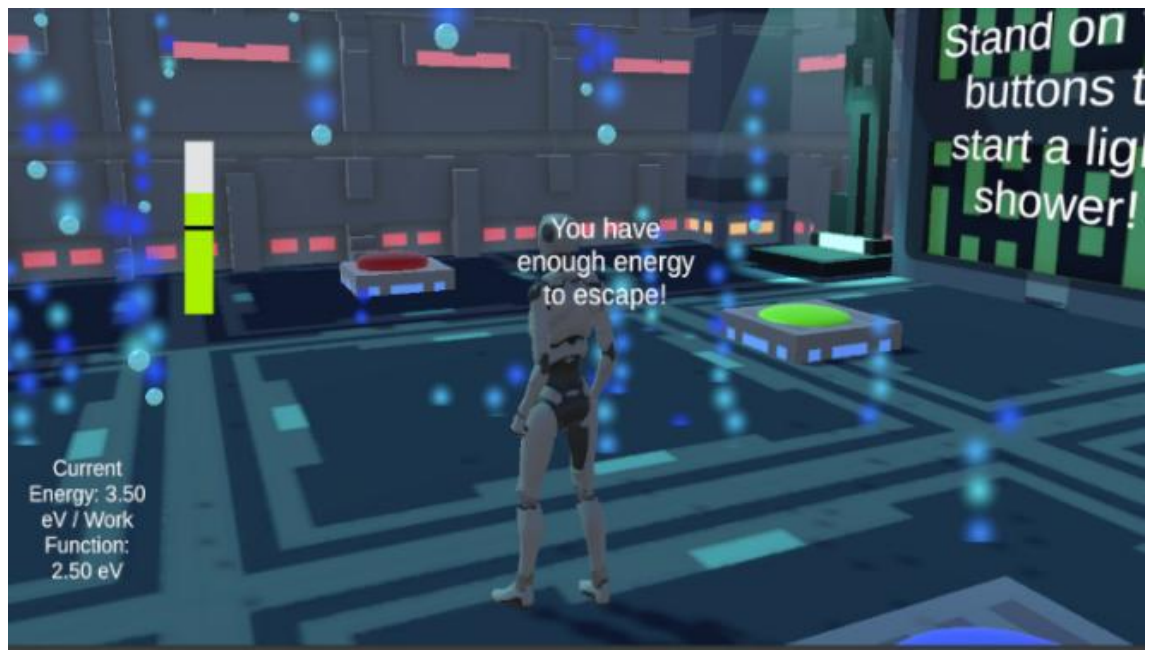

Figure 2. Photoelectric Effect Video Game

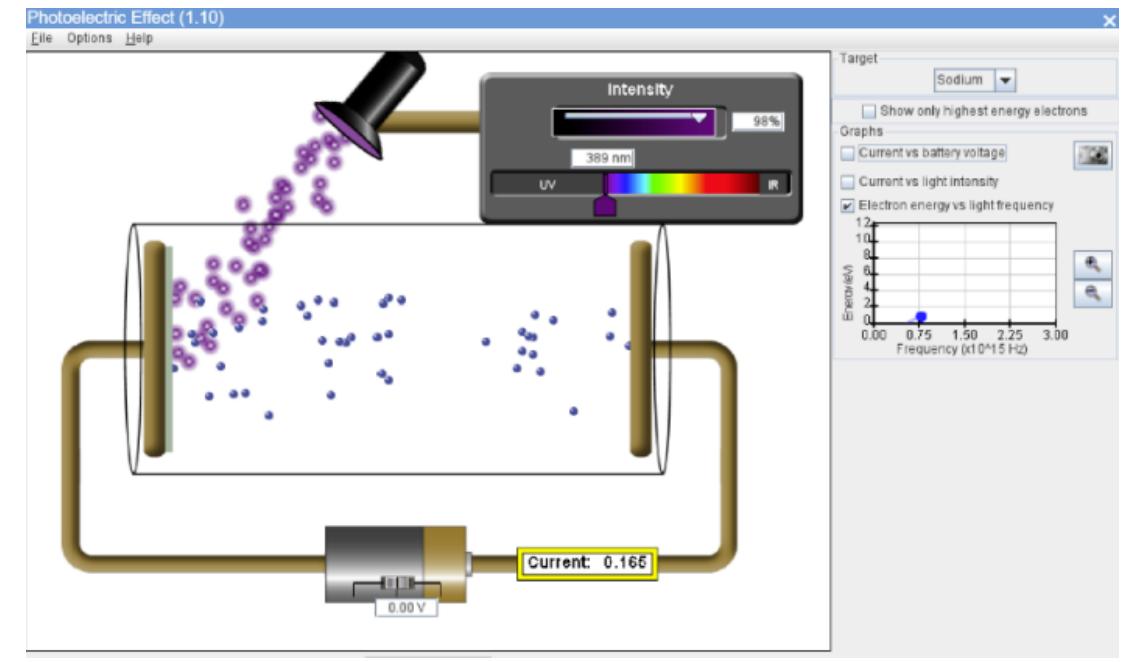

Figure 3. PhET Photoelectric Effect Simulation

**Survey :** Following both activities, students completed a reflection survey with nine open-ended questions related to their gaming habits, activity preference and challenges encountered during the game and simulation.

## 3. RESULTS

### 3.1 Correlation between Gaming Frequency and Preference

There was no significant correlation between gaming frequency and activity preference, showing that the game was accessible to learners regardless of their familiarity with gaming.

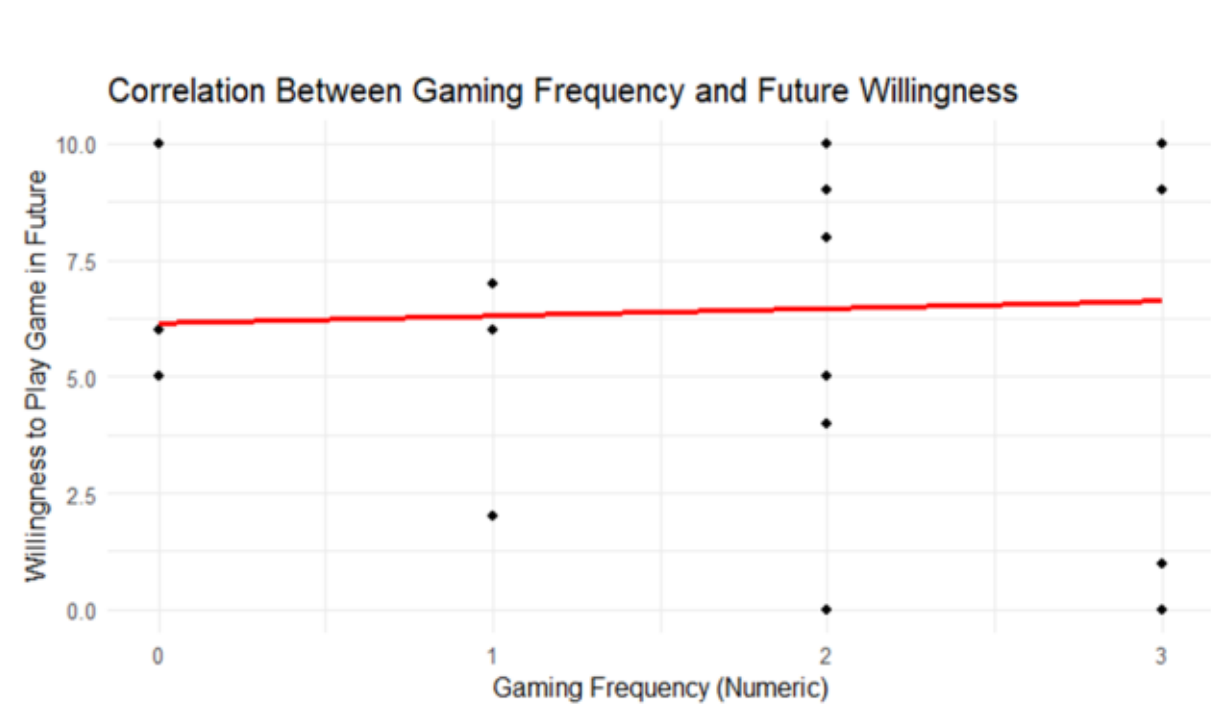

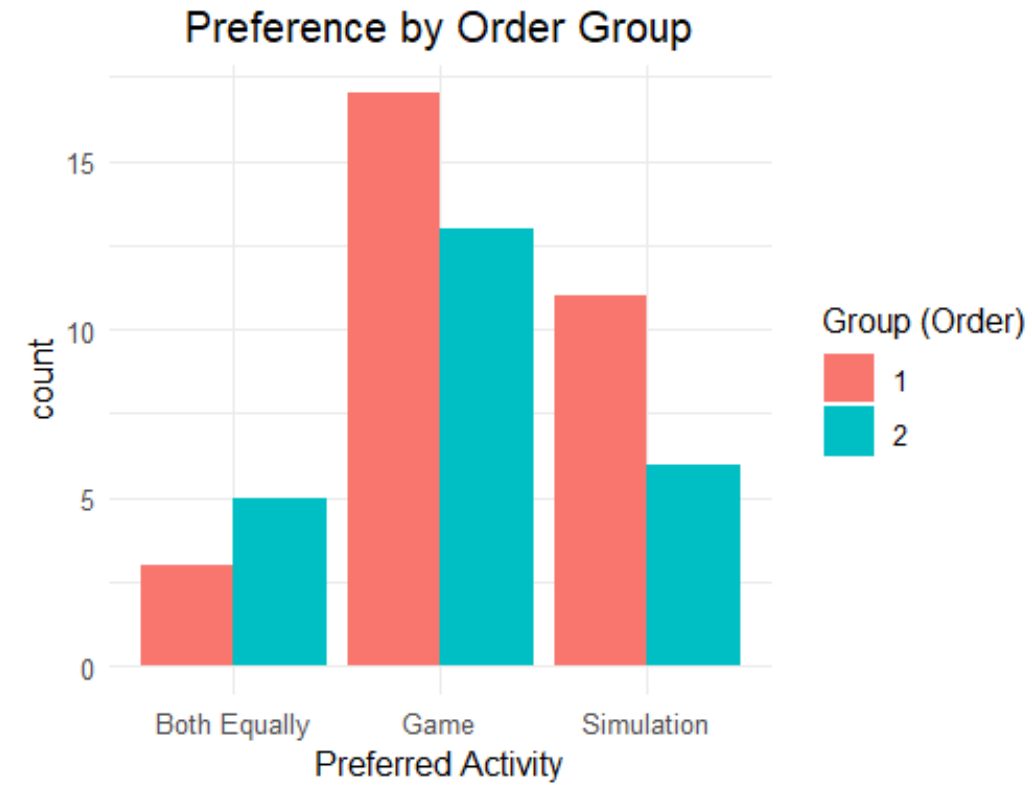

Figure 4 Correlation between Gaming Frequency and Preference

Figure 5. Preference for Activity Type

### 3.2 Preference for Activity Type

Students showed a clear preference for game-based activity (N= 30) over the simulation (N=17) (Fig. 4). The Pearson's chi squared test (p = 0.4405) showed tno significant association between preference and instructional order. Hence, the activity order did not influence students' preferences.

### 3.3 Emergent Themes

Analysis of the descriptive reflections across both groups revealed the following four overarching themes.

***Theme 1: Exploration and Intuitive Learning from Game***

The game allowed intuitive understanding of abstract, physical phenomenon by enabling students to play as the electron. By experimenting with different photon showers, they discovered which color allowed 'jumping' i.e. reaching the energy required to escape. These interactions helped students to understand concepts such as non-additivity of photon energy and concept of work function. The game's interactivity and playfulness supported engagement and curiosity, making abstract ideas feel accessible and memorable.

***Theme 2: Scientific Understanding from Simulation***

Students consistently described the simulation as a tool that enabled scientific reasoning. The electron-energy–frequency graph, the visualization of electrons leaving the metal surface, and the ability to manipulate wavelength, frequency, and intensity allowed students to directly observe the relationships central to the photoelectric effect. Many noted that the simulation *"showed the scientific side"* and supported the kinds of quantitative reasoning needed for assessments. Students valued the simulation clarity and alignment with formal physics representations, although several found the interface dense or initially overwhelming.

***Theme 3: Complementary Usefulness***

Across both groups, students perceived the game and simulation as serving distinct but complementary roles in their learning. The game was viewed as effective for introducing concepts and fostering engagement, while the simulation was more appropriate for connecting ideas to formal calculations. This complementarity appeared regardless of whether students began with the game or simulation. Students frequently suggested that using both tools together offered a more complete understanding than either one alone.

***Theme 4: Usability Challenges***

Both tools had usability issues that affected students' experiences. In the game, issues included camera movement, lag, and unclear instructions about goals or mechanics. In the simulation, students struggled with slider sensitivity, unlabeled controls, and cluttered interfaces. These issues occasionally led to confusion or misinterpretation of physical principles.

## 4. CONTRIBUTIONS AND IMPLICATIONS

This study demonstrates how educational games and simulations serve as complementary tools for teaching abstract physics concepts such as photoelectric effect. The findings show that the game supported explorative and intuitive understanding while the simulation allowed students to engage with formal representation of the concept. Further, the order of interventions did not significantly influence learning preferences. The integration of both games and simulations for teaching the photoelectric effect show how their combined use can support diverse learning styles and paces. Such integration also facilitates scaffolding, where one modality primes students for deeper engagement with the other.